# Hubble Expansion and Freeze-Out at RHIC-BES Energies from UrQMD

Gabriele Inghirami[1], Tom Reichert[2], Marcus Bleicher[2,1,3,4]

[1] *GSI Helmholtzzentrum für Schwerionenforschung GmbH, Planckstr. 1, 64291 Darmstadt, Germany*
[2] *Institut für Theoretische Physik, Goethe Universität Frankfurt,
Max-von-Laue-Strasse 1, D-60438 Frankfurt am Main, Germany*
[3] *John von Neumann-Institut für Computing, Forschungzentrum Jülich, 52425 Jülich, Germany and*
[4] *Helmholtz Forschungsakademie Hessen für FAIR (HFHF),
GSI Helmholtzzentrum für Schwerionenforschung, Campus Frankfurt,
Max-von-Laue-Str. 12, 60438 Frankfurt, Germany*

The freeze-out process in heavy ion collisions is driven by the competition between the scattering rate and the expansion rate of the matter. We analyse the expansion rate $\Theta$ (often called Hubble flow) in relativistic heavy ion collisions in the FAIR and RHIC-BES energy regimes and compare it to the scattering rate $\Gamma$ using the UrQMD transport model. We observe that the time evolution of the system is clearly separated into a compression phase and an expansion phase with time dependent $\Theta_\parallel$ and $\Theta_\perp$. The calculated values of the Hubble expansion at kinetic decoupling are in line with previous simple estimates by statistical hadronization models with a Siemens-Rasmussen type emission source. However, the actual shape of the expanding matter is, as expected, found to be between a spherically symmetric and a purely longitudinal expansion. We confirm for the first time in a microscopic simulation that the decoupling hypersurface is indeed determined by the competition of the expansion rate $\Theta$ and the scattering rate $\Gamma$ as suggested previously. This suggests that, in the range of collision energies explored in this study, simple iso-thermal/iso-energy density criteria that are often used in hybrid models to couple hydrodynamic and transport simulations may not capture the true decoupling hyper-surface.

## I. INTRODUCTION

Heavy ion collisions are often seen as a recreation of the cosmic big bang in the laboratory, hence the name 'small bang'. And indeed one observes similar features, like the expansion and the cooling of a hot and dense medium, the Quark-Gluon Plasma (QGP), with subsequent chemical and kinetic decoupling, multipole correlations of temperature fluctuation [1, 2] and Hubble flow. However, the scales are vastly different in a laboratory setting than compared to the cosmic expansion. The chemical and kinetic hadronic decoupling after the big bang are separated by $3.7 \cdot 10^5$ years, while in heavy ion collisions the two decoupling stages are separated by $1-10$ fm which is caused by and reflected in the expansion rate. At present, the Hubble constant for the "Early Universe" is estimated at $H_0 = 67.4$ km/(s Mpc) [3, 4] corresponding to $H_0 = 7.286 \cdot 10^{-36}$ fm$^{-1}$ in natural units, which is roughly 34 magnitudes smaller than the expansion rate in relativistic heavy ion collisions being on the order of 0.04 fm$^{-1}$ at HADES, GSI [5] to 0.25 fm$^{-1}$ at RHIC [6] at the decoupling stage.

At top RHIC energies the created matter has been found to behave as the most liquid substance ever observed [7, 8]. This clearly points towards the existence of QGP at high collider energies as well as a rapid thermalization. In contrast to such high energies, at the low energies investigated here there is an ongoing discussion about the onset of deconfinement and the thermal aspects of the created nuclear medium. On the one hand statistical/thermal models provide a decent description of the measured hadron yields and spectra [5, 9], however on the other hand, transport simulations show a complex dynamics driven by an intricate interplay between the compression and expansion dynamics, with time dependent shadowing of the emission due to relatively slow spectators blocking in-plane emission and thus affecting the space-time structure of the expanding emission source.

In this letter we contribute to reduce this gap by exploring the freeze-out dynamics within a microscopic simulations, however re-interpreted in quantities that allow for a straightforward and simple understanding: the freeze-out process and the corresponding hypersurface are defined by the Pomeranchuk criterion, i.e. by the interplay of the mean free paths and the characteristic size of the system [10, 11]. It can be recast as a local criterion using the expansion rate $\Theta = 1/\tau_{\rm exp}$ in comparison to the scattering rate $\Gamma_i = 1/\tau_{\rm scat,i}$ per particle of type $i$. Depending on the considered $\Gamma_i$ one can define the chemical (using $\Gamma_i$ for flavor changing reactions) and kinetic (using $\Gamma_i$ for the sum of elastic and inelastic reactions) scattering rate. Their competition defines the freeze-out hypersurface $\Sigma_\mu^{FO}$ [10–14] via the equation

$$\xi = \frac{\partial_\mu u^\mu(x)}{\sum_j \langle \sigma_{ij} v_{ij}\rangle \rho_j(x)}, \quad (1)$$

where $\sigma_{ij}$ and $v_{ij}$ are the mutual cross-sections and the relative velocities of the $i$-th and $j$-th hadron species having density $\rho_j$, while $\xi$ is a parameter with magnitude of order unity. The measurable particle yields are fixed on the chemical freeze-out hypersurface and their spectra subsequently on the kinetic freeze-out hypersurface. Extracting the hypersurface and hence the expansion and scattering dynamics directly from a microscopic transport simulation can provide several important insights:

- It allows to compare and discuss the decoupling of the system to see, if the freeze-out hypersurface is sharply defined [15].

- It allows to improve and test assumptions of the blast-wave models which usually parametrize the expansion of the system either purely longitudinally or spherically symmetric [16].

- It allows to test, if the usual assumption made in hybrid approaches for the coupling of the hydrodynamics and the Boltzmann-dynamics at fixed iso-temperature or iso-energy density are valid. Here we also refer to the discussion of the hydrodynamic decoupling in [11, 17] that indicated that the Pomeranchuk criterion yields substantially different hypersurfaces than the iso-thermal criterion.

To this aim, we explore the scattering rates and the Hubble expansion of the volume elements in central collisions of two gold nuclei at $\sqrt{s_{\mathrm{NN}}}$ = 2.4 GeV (corresponding to $E_{\mathrm{lab}}$ = 1.23 $A$GeV), $\sqrt{s_{\mathrm{NN}}}$ = 4.0 GeV and $\sqrt{s_{\mathrm{NN}}}$ = 7.7 GeV. For our study we employ the UrQMD transport approach supplemented by a coarse graining method to extract the local flow velocities and the scattering rates. This approach has demonstrated its applicability for a wide range of investigations ranging from dilepton [18, 19] and photon [20] studies to charm dynamics [21] and transport coefficients [22]. For related studies on the chemical and kinetic freeze-out we refer to [23, 24]. Here we use the same approach to investigate the Hubble expansion and the scattering dynamics by microscopic coarse grained simulations. We then compare to typical estimates of the Hubble flow and to previous results from hydrodynamic simulations.

## II. THE URQMD MODEL AND THE COARSE GRAINING APPROACH

We employ the Ultra relativistic Quantum Molecular Dynamics (UrQMD) transport model in version v3.4 [25, 26] in cascade mode. UrQMD is based on the covariant propagation of hadrons and their interactions by elastic and/or inelastic collisions. Relevant cross sections are taken, if available, from experimental data or derived from effective models. UrQMD includes mesonic and baryonic resonances up to masses of 2 GeV and has a longstanding history to reproduce hadron yields and spectra and forecast new phenomena. In case of the FAIR and RHIC-BES energy regimes investigated in the present study, UrQMD has been well tested and successfully describes the dynamics of the collision. For recent results in this energy regime, we refer the reader to [27–29].

The UrQMD coarse-graining approach [18–24, 30] in general consists in computing the temperature and the baryon chemical potential from the average energy-momentum tensor and net baryon current of the hadrons formed in a large set of heavy ion collision events with the same collision energy and centrality. The computation is done in the cells of a fixed spatial grid at constant intervals of time. In the present study, the cells are four-cubes with spatial sides of length $\Delta x_i$ = 1 fm and $\Delta t = 0.25$ fm length in time direction. In this article, the coarse-graining approach is deployed to calculate the four-divergence of the four-velocity field. First, we evaluate the net-baryon four current $j_{\mathrm{B}}^{\mu}$ as

$$j_{\mathrm{B}}^{\mu}(t,\vec{r}) = \frac{1}{V} \left\langle \sum_{i=1}^{N_h \in V} B_i \frac{p_i^{\mu}}{p_i^0} \right\rangle, \qquad (2)$$

in which $V$ is the volume of the cell, $B_i$ is the baryon number and $p_i^{\mu}$ stands for four momentum of the hadron $i$. The sums run over all hadrons $N_h$ in the cell. We adopt the Eckart's frame definition [31], identifying the fluid frame with that of the net baryon current. We obtain the fluid four velocity $u^{\mu}$ from $j_{\mathrm{B}}^{\mu}$ as $u^{\mu} = j_{\mathrm{B}}^{\mu}/|j_{\mathrm{B}}| = \gamma(1,\vec{v})$, in which $\gamma$ is the Lorentz factor and $\vec{v}$ the fluid velocity. We then compute the four-divergence $\partial_{\mu}u^{\mu}$ to obtain the expansion (Hubble) coefficient.

## III. RESULTS

All calculations are done for Au+Au reactions at impact parameter $b = 0$ fm. The fluid velocity always refers to the center of momentum frame, however note that the expansion scalar $\Theta = \partial_{\mu}u^{\mu}(x)$ is of course independent of the frame. We further adopt the following shorthand notations throughout the rest of the manuscript to discuss the different components of the expansion rate (frame and coordinates dependent):

$$\Theta_{\perp} \equiv \frac{\partial \gamma}{\partial t} + \frac{\partial u^x}{\partial x} + \frac{\partial u^y}{\partial y}, \qquad (3)$$

$$\Theta_{\parallel} \equiv \frac{\partial \gamma}{\partial t} + \frac{\partial u^z}{\partial z}, \qquad (4)$$

$$\Theta_x \equiv \frac{\partial \gamma}{\partial t} + \frac{\partial u^x}{\partial x}, \qquad (5)$$

where $\gamma$ is the Lorentz gamma factor of the fluid velocity $u^{\mu} = \gamma(1, v^x, v^y, v^z)$.

Let us start our investigation with the time dependence of the expansion rate. Fig. 1 shows the time dependence of the (Hubble) expansion rate in the transverse ($\Theta_{\perp}$) and longitudinal direction ($\Theta_{\parallel}$) in the center of system (i.e. in a cube with side length 1 fm, centered at $x = y = z = 0$) for Au+Au collisions at $\sqrt{s_{\mathrm{NN}}}$ = 2.4 GeV (corresponding to $E_{\mathrm{lab}}$ = 1.23 $A$GeV, red line), $\sqrt{s_{\mathrm{NN}}}$ = 4.0 GeV (yellow line) and $\sqrt{s_{\mathrm{NN}}}$ = 7.7 GeV (green line) from UrQMD. We remark that in this cell the average velocity vanishes due to symmetry, but not the derivatives of the velocities[1]. The simulation results are compared with the fits to their tails following the function $k/(\tau - \tau_0)$. We recall that in a system initiating its expansion at $\tau_0 = 0$, one expects that in the case of a purely longitudinal expansion $k = 1$ and in case of a spherically symmetric expanding system that $k = 3$ [33]. Given the strong anisotropy of the system, we study separately the longitudinal and the radial expansion. We find that $k_\parallel \approx 1.15-1.33$ for the $\Theta_\parallel$ and $k_\perp \approx 2.23-2.33$ for $\Theta_\perp$ allows for a good fit of the time evolution of the system after the equilibration phase. These fitted values remain within a narrow range despite a factor 3 increase in the collision energy. This suggests that the expansion can be well described by a Hubble-like behavior. On the other hand, in line with the expectations, we also observe that the time evolution of the system is dissected into two stages: firstly a compression phase in the longitudinal direction and secondly an expansion phase (in both the longitudinal and radial directions) after full overlap is achieved. The dynamics in the compression and system equilibration phase strongly depends on the collision energy, with characteristic times that vary broadly, as shown by the time displacements of the maxima in the plots and as reflected by the different values of the fitted $\tau_0$ parameter.

---

[1] We compute the derivatives with Numpy [32] using a standard second order algorithm that exploits the adjacent cells.

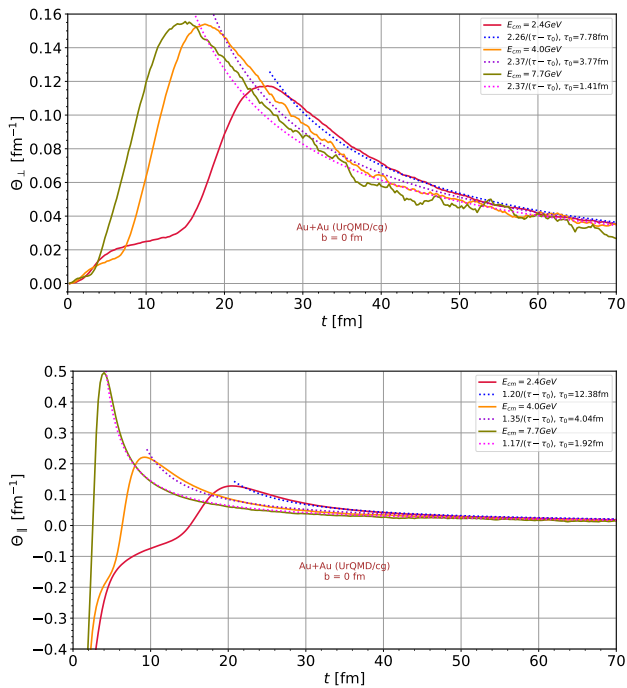

FIG. 1. [Color online] Calculated time dependence of the expansion rate $\Theta$ in the transverse ($\Theta_\perp$) and longitudinal direction ($\Theta_\parallel$) in the center of the system ($x = y = z = 0$) for Au+Au reactions at at $\sqrt{s_{NN}} = 2.4$ (red), 4.0 (yellow) and 7.7 GeV (green). The simulation results are shown as full lines, while the fits to the function $k/(\tau - \tau_0)$ are shown as dotted lines.

Fig. 2 shows the space dependence of $\Theta_x$, the ex-pansion rate in the transverse direction, at $y = z = 0$, for $t = 30$ fm and $t = 60$ fm from UrQMD simulations for Au+Au collisions at $\sqrt{s_{NN}} = 2.4$, 4.0 and 7.7 GeV. The simulation results are compared with $\Theta = 1/\tau$, with $\tau = \sqrt{t^2 - z^2}$. Since we are simulating perfectly central collisions (impact parameter $b = 0$ fm), apart from the statistical fluctuations in the positions of the nucleons forming the colliding nuclei and in their Fermi momenta, our system has a radial symmetry with respect to the beam ($z$) axis. We observe that at the lowest energies, the calculated radial expansion can only approximately be described by a simple Hubble flow profile (dotted line), but has a more complicated structure indicating a de-layed expansion as compared to the simple estimates. At higher energies, the system approaches the expectations from the free hydrodynamic expansion and starts to follow the Hubble flow picture well.

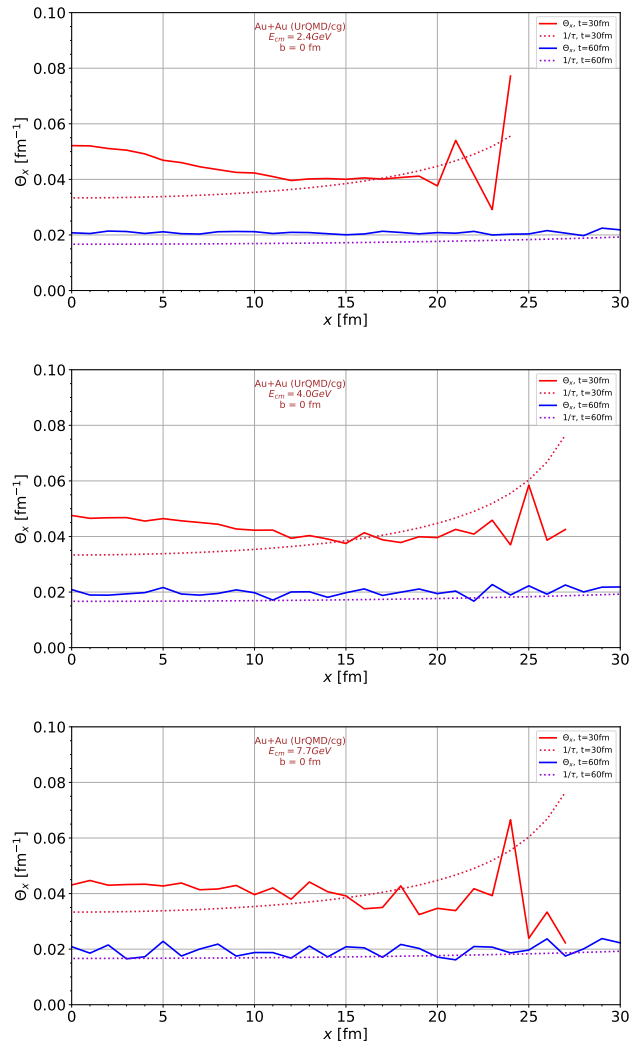

FIG. 2. [Color online] Calculated space dependence of $\Theta_x$, the Hubble expansion rate in the transverse direction in the center of the system ($z = y = 0$) for $t = 30$ fm and $t = 60$ fm for Au+Au reactions at $\sqrt{s_{NN}} = 2.4$ (top), 4.0 (middle) and 7.7 GeV (bottom). The simulation results (full lines) are compared with $\Theta = 1/\tau$, with $\tau = \sqrt{t^2 - x^2}$ (dotted lines). For clarity, we removed the statistical fluctuations at the shock front corresponding to $x \approx 30$ fm.

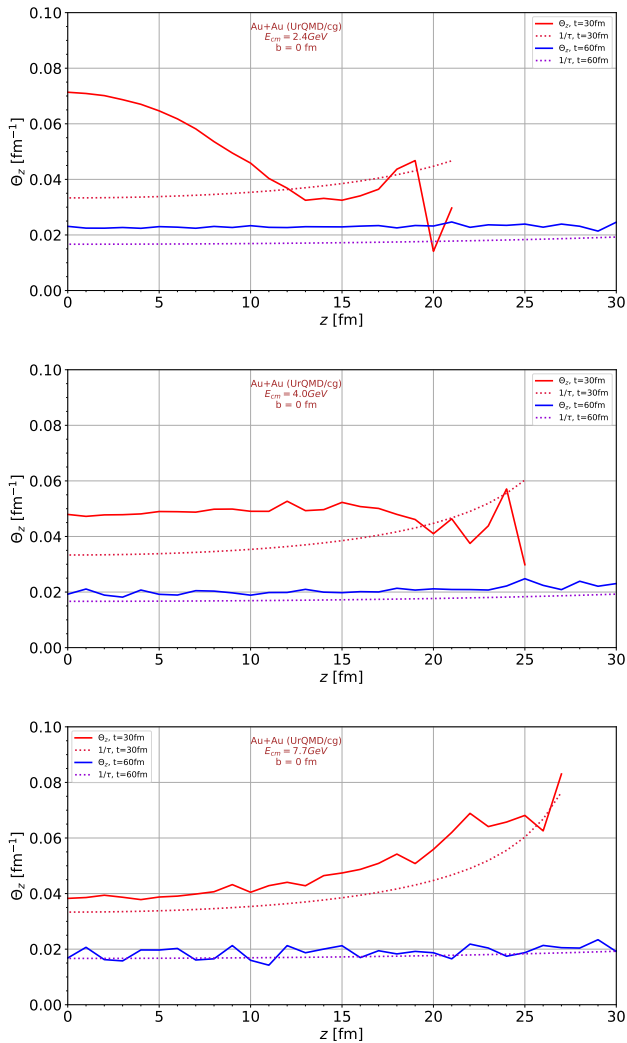

FIG. 3. [Color online] Calculated space dependence of $\Theta_z$, the expansion rate in the longitudinal direction in the center of the system ($x = y = 0$) for $t = 30$ fm and $t = 60$ fm for Au+Au reactions at $\sqrt{s_{NN}} = 2.4$ (top), 4.0 (middle) and 7.7 GeV (bottom). The simulation results (full lines) are compared with $\Theta = 1/\tau$, with $\tau = \sqrt{t^2 - z^2}$ (dotted lines). For clarity, we removed the statistical fluctuations at the shock front corresponding to $z \approx 30$ fm.

Fig. 3 shows the space dependence of the expansion rate $\Theta$ in the longitudinal direction ($\Theta_\parallel$) in the center of the system ($x = y = 0$) for $t = 30$ fm and $t = 60$ fm from UrQMD at $\sqrt{s_{NN}} = 2.4$, 4.0 and 7.7 GeV. The simulation results are compared with $\Theta = 1/\tau$, as it would be expected in a purely longitudinal expansion, with $\tau = \sqrt{t^2 - z^2}$. Here we observe a similar feature as for the transverse expansion: at the lowest energy, the longitudinal expansion is delayed and starts from the center (the region with the highest pressure) of the collision. Towards higher energies, the simulation tends to follow the simple Hubble flow profile expected for a strongly expanding system.

After the extraction of the expansion rate we can test the validity of the assumption of the blast wave model which is often used by experimental groups to interpret their results. A central ingredient of the blast wave model is the radial expansion velocity and the form of the radial velocity profile. Within our approach we are able to test the standard assumptions of the blast wave model by calculating $v(r_\perp)$ from the transport simulation at different times and compare to the blast wave parametrization $v(r_\perp) = \tanh(\Theta_{BW} r_\perp)$ [5]. Moreover, we perform a least squares fit to extract a numerical value for $\Theta(t)$ from our calculation. The results are shown in Fig. 4. We find that the usual blast wave flow profile indeed provides a reasonable description of the observed radial velocity profile. In addition we confirm that the blast wave parameter $\Theta_{BW}$ indeed corresponds to the transverse expansion rate.

In Table I we display the results of the least squares fits, based in one case on the $v(r_\perp)$ obtained from our simulation and in the other case on $\Theta$ computed from the simulation. We notice an overall fair consistency within the results, with equal time values of $\Theta(t)$ at different collision energies quantitatively close to each other.

| $\sqrt{s_{NN}}$ (GeV) | $t$ (fm) | $\Theta(v(r_\perp))$ | $\Theta_{comp.}$ |
|---|---|---|---|
| 2.4 | 20 | 0.0595 | 0.0646 |
| 2.4 | 40 | 0.0353 | 0.0313 |
| 2.4 | 60 | 0.0216 | 0.0208 |
| 4.0 | 20 | 0.0657 | 0.0609 |
| 4.0 | 40 | 0.0340 | 0.0308 |
| 4.0 | 60 | 0.0205 | 0.0203 |
| 7.7 | 20 | 0.0620 | 0.0574 |
| 7.7 | 40 | 0.0326 | 0.0290 |
| 7.7 | 60 | 0.0199 | 0.0200 |

TABLE I. Least squares fitted values of $\Theta$ at various times and for various collision energies. In the left column we used the equation $v(r_\perp) = \tanh(\Theta(t) r_\perp)$, with $v(r_\perp)$ known from our coarse graining data and $\Theta(t)$ unknown, while in the right column we used only the RHS of the previous equation and we fitted the values of $\Theta(r, t)$ computed from the coarse graining data.

Let us finally come to the freeze-out and decoupling of the system. Here we relate our studies to the works of Shuryak and Huovinen [11, 17] that have already pointed out, that the freeze-out of the system should be described by the interplay of expansion and scattering rates. Given the UrQMD dynamics we are now in the position to connect the kinetic decoupling to the local scattering and expansion rate and test this hypothesis. The upper part of Fig. 5 shows the the evolution of the ratio between the expansion rate $\Theta$ and scattering rate per hadron $\Gamma$ for Au+Au collisions at $\sqrt{s_{NN}} = 2.4$, 4.0 and 7.7 GeV. We clearly observe an initial phase dominated by the scatter-ing rate, relatively short at $\sqrt{s_{\rm NN}} = 7.7$ GeV, but lasting until 6 fm and 15 fm for $\sqrt{s_{\rm NN}} = 4.0$ GeV and $\sqrt{s_{\rm NN}} = 2.4$ GeV, respectively. Afterwards, the expansion rate starts to grow and eventually dominates, as it is expected in a system that becomes more and more diluted. The lower part of Fig. 5 shows the normalized kinetic freeze-out distributions. While the distribution of the freeze-out rates is rather broad it can still be seen that the peak emission corresponds to a well defined value of $\Theta_\parallel/\Gamma = 0.5$ (independent of energy) for the matter in the central cell. Thus, we confirm the idea [11] that the freeze-out is indeed driven and defined by the competition of the local scattering rates versus the expansion rates.

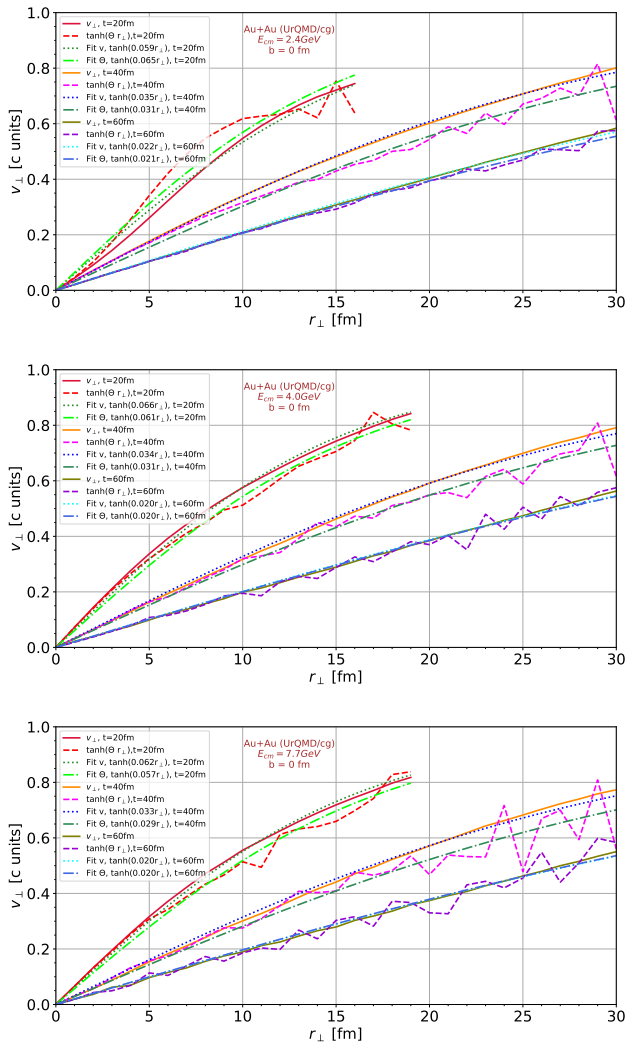

FIG. 4. [Color online] Radial flow profiles (at $y = z = 0$) at various fixed times in comparison to the fits using $v(r_\perp) = \tanh(\Theta(t) r_\perp)$ for $t = 20$ fm, 40 fm and 60 fm for Au+Au reactions at $\sqrt{s_{\rm NN}} = 2.4$ (top), 4.0 (middle) and 7.7 GeV (bottom). For clarity, we removed the oscillations at the shock front.

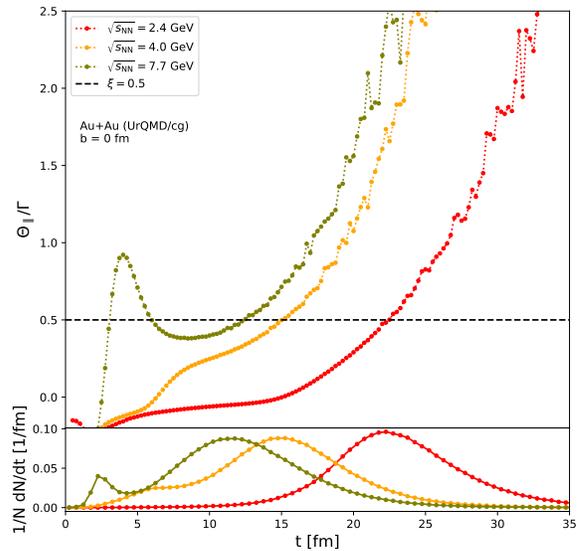

FIG. 5. [Color online] Upper part: time evolution of the expansion rate and scattering rate in the central cube for Au+Au reactions at $\sqrt{s_{\rm NN}} = 2.4$ (red), 4.0 (yellow) and 7.7 GeV (green). Lower part: normalized time distribution of the hadron kinetic freeze-out rate in the central cube. The figure refers to Au+Au collisions at $\sqrt{s_{\rm NN}} = 2.4$, 4.0 and 7.7 GeV.

## IV. SUMMARY

In this article we have analyzed the expansion rate and the scattering rate in relativistic heavy ion collisions using a coarse graining approach based on numerical transport simulations with UrQMD. We focused on central Au+Au collisions at $\sqrt{s_{\rm NN}} = 2.4$, 4.0 and 7.7 GeV, corresponding to HADES, FAIR and low end RHIC BES energies. Given the anisotropy of the system, we considered separately the transverse and the longitudinal direction, the latter parallel to the beam axis. We showed that, after an initial compression and equilibration phase, the average evolution of the longitudinal and transverse expansion rate in the center of the system shows a similar expansion flow at all investigated collision energies and is well captured by a Hubble-like expansion. We evaluated the transverse and longitudinal dependence of the expansion rate at two different times, showing that at $\tau = 30$ fm the system still exhibits a non uniform behavior, while at $\tau = 60$ fm it can be effectively modeled with the simple $1/\tau$ law, with a degree of approximation that improves with increasing collision energy. We studied the profile of the radial velocity $v(r_\perp)$ at t = 20, 40 and 60 fm and we compared it with the $\tanh(\Theta(t) r_\perp)$ relation often used in blast wave models. We showed that the blast wave assumptions are justified in comparison to the full dynamics modelled here and validated that the blast wave model can be used to extract a meaningful Hubble parameter for the late stage of the reaction. Finally, we evaluated the ratio $\Theta_\parallel/\Gamma$ between the expansion and the scattering rate per baryon, comparing it with the normalized time distribution of the kinetic freeze-out. We found a clear correspondence between the maximum of the kinetic freeze-out rate and $\Theta_\parallel/\Gamma = 0.5$ at all investigated energies.

## ACKNOWLEDGMENTS

This work was supported in the framework of COST Action CA15213 THOR and we acknowledge support from STRONG-2020, financed by the Horizon 2020 program of the European Community. Computational resources were provided by the Center for Scientific Computing (CSC) of the Goethe University. G. Inghirami acknowledges funding by the Deutsche Forschungsgemeinschaft (DFG, German Research Foundation) – Project number 315477589 – TRR 211.

## REFERENCES

[1] P. A. R. Ade *et al.* [Planck], Astron. Astrophys. **571**, A15 (2014) doi:10.1051/0004-6361/201321573 [arXiv:1303.5075 [astro-ph.CO]].
[2] B. Schenke, P. Tribedy and R. Venugopalan, Phys. Rev. Lett. **108**, 252301 (2012) doi:10.1103/PhysRevLett.108.252301 [arXiv:1202.6646 [nucl-th]].
[3] N. Aghanim *et al.* [Planck], Astron. Astrophys. **641**, A6 (2020) doi:10.1051/0004-6361/201833910 [arXiv:1807.06209 [astro-ph.CO]].
[4] L. Verde, T. Treu and A. G. Riess, Nature Astron. **3**, 891 doi:10.1038/s41550-019-0902-0 [arXiv:1907.10625 [astro-ph.CO]].
[5] S. Harabasz, W. Florkowski, T. Galatyuk, ‡. Ma Lgorzata Gumberidze, R. Ryblewski, P. Salabura and J. Stroth, Phys. Rev. C **102**, no.5, 054903 (2020) doi:10.1103/PhysRevC.102.054903 [arXiv:2003.12992 [nucl-th]].
[6] M. Chojnacki, W. Florkowski and T. Csorgo, Phys. Rev. C **71**, 044902 (2005) doi:10.1103/PhysRevC.71.044902 [arXiv:nucl-th/0410036 [nucl-th]].
[7] P. Romatschke and U. Romatschke, Phys. Rev. Lett. **99**, 172301 (2007) doi:10.1103/PhysRevLett.99.172301 [arXiv:0706.1522 [nucl-th]].
[8] H. Song, S. A. Bass, U. Heinz, T. Hirano and C. Shen, Phys. Rev. Lett. **106**, 192301 (2011) [erratum: Phys. Rev. Lett. **109**, 139904 (2012)] doi:10.1103/PhysRevLett.106.192301 [arXiv:1011.2783 [nucl-th]].
[9] A. Andronic, P. Braun-Munzinger and J. Stachel, Nucl. Phys. A **772**, 167-199 (2006) doi:10.1016/j.nuclphysa.2006.03.012 [arXiv:nucl-th/0511071 [nucl-th]].
[10] J. P. Bondorf, S. I. A. Garpman and J. Zimanyi, Nucl. Phys. A **296**, 320-332 (1978) doi:10.1016/0375-9474(78)90076-3
[11] C. M. Hung and E. V. Shuryak, Phys. Rev. C **57**, 1891-1906 (1998) doi:10.1103/PhysRevC.57.1891 [arXiv:hep-ph/9709264 [hep-ph]].
[12] K. J. Eskola, H. Niemi and P. V. Ruuskanen, Phys. Rev. C **77**, 044907 (2008) doi:10.1103/PhysRevC.77.044907 [arXiv:0710.4476 [hep-ph]].
[13] Y. M. Sinyukov, S. V. Akkelin, I. A. Karpenko and Y. Hama, Acta Phys. Polon. B **40**, 1025-1036 (2009) [arXiv:0901.1576 [nucl-th]].
[14] U. W. Heinz, Landolt-Bornstein **23**, 240 (2010) doi:10.1007/978-3-642-01539-7_9 [arXiv:0901.4355 [nucl-th]].
[15] J. Knoll, Nucl. Phys. A **821**, 235-250 (2009) doi:10.1016/j.nuclphysa.2009.01.079 [arXiv:0803.2343 [nucl-th]].
[16] F. Retiere and M. A. Lisa, Phys. Rev. C **70**, 044907 (2004) doi:10.1103/PhysRevC.70.044907 [arXiv:nucl-th/0312024 [nucl-th]].
[17] H. Holopainen and P. Huovinen, J. Phys. Conf. Ser. **509**, 012114 (2014) doi:10.1088/1742-6596/509/1/012114 [arXiv:1310.0347 [nucl-th]].
[18] P. Huovinen, M. Belkacem, P. J. Ellis and J. I. Kapusta, Phys. Rev. C **66**, 014903 (2002) doi:10.1103/PhysRevC.66.014903 [arXiv:nucl-th/0203023 [nucl-th]].
[19] S. Endres, H. van Hees, J. Weil and M. Bleicher, Phys. Rev. C **92**, no.1, 014911 (2015) doi:10.1103/PhysRevC.92.014911 [arXiv:1505.06131 [nucl-th]].
[20] S. Endres, H. van Hees and M. Bleicher, Phys. Rev. C **93**, no.5, 054901 (2016) doi:10.1103/PhysRevC.93.054901 [arXiv:1512.06549 [nucl-th]].
[21] G. Inghirami, H. van Hees, S. Endres, J. M. Torres-Rincon and M. Bleicher, Eur. Phys. J. C **79**, no.1, 52 (2019) doi:10.1140/epjc/s10052-019-6537-6 [arXiv:1804.07751 [hep-ph]].
[22] T. Reichert, G. Inghirami and M. Bleicher, Phys. Lett. B **817**, 136285 (2021) doi:10.1016/j.physletb.2021.136285 [arXiv:2011.04546 [nucl-th]].
[23] G. Inghirami, P. Hillmann, B. Tomášik and M. Bleicher, J. Phys. G **47**, no.2, 025104 (2020) doi:10.1088/1361-6471/ab53f4 [arXiv:1909.00643 [hep-ph]].
[24] T. Reichert, G. Inghirami and M. Bleicher, Eur. Phys. J. A **56**, no.10, 267 (2020) doi:10.1140/epja/s10050-020-00273-y [arXiv:2007.06440 [nucl-th]].
[25] S. A. Bass, M. Belkacem, M. Bleicher, M. Brandstetter, L. Bravina, C. Ernst, L. Gerland, M. Hofmann, S. Hofmann and J. Konopka, *et al.* Prog. Part. Nucl. Phys. **41**, 255-369 (1998) doi:10.1016/S0146-6410(98)00058-1 [arXiv:nucl-th/9803035 [nucl-th]].
[26] M. Bleicher, E. Zabrodin, C. Spieles, S. A. Bass, C. Ernst, S. Soff, L. Bravina, M. Belkacem, H. Weber and H. Stoecker, *et al.* J. Phys. G **25**, 1859-1896 (1999) doi:10.1088/0954-3899/25/9/308 [arXiv:hep-ph/9909407 [hep-ph]].
[27] S. Sombun, K. Tomuang, A. Limphirat, P. Hillmann, C. Herold, J. Steinheimer, Y. Yan and M. Bleicher, Phys. Rev. C **99**, no.1, 014901 (2019) doi:10.1103/PhysRevC.99.014901 [arXiv:1805.11509 [nucl-th]].
[28] P. Hillmann, J. Steinheimer, T. Reichert, V. Gaebel, M. Bleicher, S. Sombun, C. Herold and A. Limphirat, J. Phys. G **47**, no.5, 055101 (2020) doi:10.1088/1361-6471/ab6fcf [arXiv:1907.04571 [nucl-th]].

[29] T. Reichert, P. Hillmann and M. Bleicher, Nucl. Phys. A **1007**, 122058 (2021) doi:10.1016/j.nuclphysa.2020.122058 [arXiv:2004.10539 [nucl-th]].
[30] S. Endres, H. van Hees, J. Weil and M. Bleicher, Phys. Rev. C **91**, no.5, 054911 (2015) doi:10.1103/PhysRevC.91.054911 [arXiv:1412.1965 [nucl-th]].
[31] C. Eckart, Phys. Rev. **58**, 919-924 (1940) doi:10.1103/PhysRev.58.919
[32] C. R. Harris, K. J. Millman, S. J. van der Walt, R. Gommers, P. Virtanen, D. Cournapeau, E. Wieser, J. Taylor, S. Berg and N. J. Smith, *et al.* Nature **585**, no.7825, 357-362 (2020) doi:10.1038/s41586-020-2649-2 [arXiv:2006.10256 [cs.MS]].
[33] A. Dumitru, Phys. Lett. B **463**, 138 (1999) doi:10.1016/S0370-2693(99)01018-7 [arXiv:hep-ph/9905217 [hep-ph]].